\documentclass[12pt]{article} 
\usepackage[xdvi]{graphicx} 
\usepackage{amssymb}
\usepackage{amsmath} 
 
\begin{document}   
\newcommand{\todo}[1]{{\em \small {#1}}\marginpar{$\Longleftarrow$}}   
\newcommand{\labell}[1]{\label{#1}\qquad_{#1}} 

\rightline{DCPT/01/49}   
\rightline{hep-th/0106220}   
\vskip 1cm


\begin{center} 
{\Large \bf Stability and the negative mode for Schwarzschild in a
finite cavity}
\end{center} 
\vskip 1cm   
  
\renewcommand{\thefootnote}{\fnsymbol{footnote}}   
\centerline{\bf   
 James P. Gregory\footnote{J.P.Gregory@durham.ac.uk} and Simon 
F. Ross\footnote{S.F.Ross@durham.ac.uk}}    
\vskip .5cm   
\centerline{ \it Centre for Particle Theory, Department of  
Mathematical Sciences}   
\centerline{\it University of Durham, South Road, Durham DH1 3LE, U.K.}   
  
\setcounter{footnote}{0}   
\renewcommand{\thefootnote}{\arabic{footnote}}


\begin{abstract}   
It has been proposed that translationally-invariant black branes are
classically stable if and only if they are locally thermodynamically
stable. Reall has outlined a general argument to demonstrate this,
and studied in detail the case of charged black p-branes in type II
supergravity \cite{Reall:branes}. We consider the application of his
argument in the simplest non-trivial case, an uncharged asymptotically
flat brane  enclosed in a finite cylindrical cavity. In this
simple context, it is possible to give a more complete argument than
in the cases considered earlier, and it is therefore a particularly
attractive example of the general approach.
\end{abstract}

\section{Introduction}     

It was recently proposed, in a conjecture of Gubser and
Mitra~\cite{Gubser:instability1,Gubser:instability2}, that a black
brane with a non-compact translational symmetry is classically stable
if, and only if, it is locally thermodynamically stable.  In
subsequent work~\cite{Reall:branes}, Reall provided a general argument
for the validity of this conjecture, based on establishing a
relationship between the classical Gregory-Laflamme
instability~\cite{Gregory:instability1,Gregory:instability2} and a
Euclidean negative mode associated with thermodynamic instability.
Interesting tests of this conjecture are provided by solutions which
pass from (classical or thermodynamic) stability to instability as
some parameter is varied. In~\cite{Reall:branes}, charged black
p-brane solutions were considered, where in some cases the specific
heat becomes positive near extremality, and it was explicitly
demonstrated that for suitable gauge choices, the equations satisfied
by the Euclidean negative mode and the classical Gregory-Laflamme
perturbation are equivalent.

In this paper, we will consider the simpler case of an uncharged black
$p$-brane in $d+p$ dimensions, constructed by taking a direct product
of a $d$-dimensional Schwarzschild black hole and $p$ flat
directions. We consider this system in a finite cavity, so we can vary
the specific heat by varying the ratio of the cavity size and the
black hole's mass. The relation between the classically unstable mode and
the Euclidean negative mode for this solution was already considered
in infinite volume in~\cite{Reall:branes}. We show that the
introduction of the cavity walls does not spoil the equivalence, and
furthermore show that the negative mode coincides precisely with the
regime of thermodynamic instability.\footnote{A similar calculation of
the negative mode for Euclidean Schwarzschild-anti de Sitter was
carried out in~\cite{Prestidge:stability}. It was also shown
in~\cite{timsthesis} that for Reissner-Nordstr\"om black holes in
infinite volume, the negative mode coincides with the regime of
thermodynamic instability.}

To begin, we briefly review the discussion in~\cite{Reall:branes}. In
the semi-classical approximation to the Euclidean path integral
\begin{equation}
Z = \int D[g] e^{-I[g]},
\end{equation}
we write the metric as $g_{\mu\nu}=\bar{g}_{\mu\nu}+h_{\mu\nu}$, where
$\bar{g}_{\mu\nu}$ is some Euclidean solution of the field equations and
$h_{\mu\nu}$ is a small perturbation. The action can then be approximated
by
\begin{equation}
I_E[g]=I_E[\bar{g}]+\int d^4x
\sqrt{\bar{g}}h^{\mu\nu}A_{\mu\nu\rho\sigma}h^{\rho\sigma} .
\end{equation} 
If the operator $A_{\mu\nu\rho\sigma}$ in the quadratic term has
negative eigenvalues, there will be an imaginary part in the partition
function, and the classical solution $\bar{g}_{\mu\nu}$ is interpreted
as a saddle-point. If we decompose the perturbation into a transverse
tracefree part $h^{TT}_{\mu\nu}$, a trace and a longitudinal part, the
trace will have a negative quadratic term, but this is just a sign of
the usual conformal factor problem, and the resulting imaginary part
is cancelled by a corresponding contribution from integrating over
ghosts, so it does not correspond to a physical
instability~\cite{Gross:flatspace}. The quadratic term for the
longitudinal part is positive definite. The physical negative modes in
pure gravity therefore arise only from the quadratic term for
$h^{TT}_{\mu\nu}$, which involves the Euclidean Lichnerowicz operator
\begin{equation} 
G_{\mu\nu\rho\sigma}=-\bar{g}_{\mu\rho}\bar{g}_{\nu\sigma}\nabla_{\beta}
\nabla^{\beta}-2\bar{R}_{\mu\rho\nu\sigma}.   
\end{equation} 
The negative modes are given by the eigenvectors of 
\begin{equation} \label{eqn:eigenvalue} 
G_{\mu\nu}{}^{\rho\sigma} h^{TT}_{\rho\sigma} = \lambda h^{TT}_{\mu\nu} 
\end{equation}
with negative eigenvalues. 

In~\cite{Prestidge:stability}, it was shown that Schwarzschild black
holes with negative specific heat will necessarily have a negative
mode. This was extended to the charged black brane solutions
in~\cite{Reall:branes}. Thus, thermodynamic instability in the
canonical ensemble implies the existence of a negative mode. However,
no general proof of the converse exists. In this paper we will check 
this explicitly by finding the lowest eigenvalue of
(\ref{eqn:eigenvalue}) and showing it is negative if and only if the
specific heat is negative. 

The key remaining step in relating thermodynamic instability and
classical instability is relating the negative mode satisfying
(\ref{eqn:eigenvalue}) in the black hole solution to the unstable mode
in the Gregory-Laflamme analysis of the perturbations about a black
brane solution. For the uncharged black branes we are interested in,
this connection is very direct~\cite{Reall:branes}. The metric for the
brane is
\begin{equation} 
ds^2 = g_{\mu\nu}dx^{\mu}dx^{\nu} + \delta_{ij}dz^{i}dz^{j}, 
\end{equation} 
where $g_{\mu\nu}$ is the $d$-dimensional Lorentzian Schwarzschild 
metric and $z^{i}$ are $p$ flat spatial directions. If we consider a
metric perturbation
\begin{equation} 
h_{\mu\nu}=\exp\left(i\mu_{i}z^{i}\right)H_{\mu\nu},\qquad h_{\mu 
i}=h_{ij}=0, 
\end{equation} 
where $H_{\mu\nu}$ is transverse traceless and
spherically symmetric, then the equation of motion
implies~\cite{Gregory:instability1,Gregory:instability2}  
\begin{equation} \label{eqn:GL}
\tilde{G}_{\mu\nu}{}^{\rho\sigma} H_{\rho\sigma} = -\mu^2 H_{\mu\nu}, 
\end{equation}
where $\tilde{G}$ is now the Lorentzian Lichnerowicz operator for the
metric $g_{\mu\nu}$. A time-independent solution of this equation with
non-zero $\mu$ represents a threshold unstable mode, separating stable
and unstable perturbations. But on time-independent perturbations,
(\ref{eqn:eigenvalue}) and (\ref{eqn:GL}) are equivalent.  Thus the
conditions for classical instability and existence of a Euclidean
negative mode give rise to the same equation.

The introduction of a finite boundary introduces little modification
in this argument connecting the Euclidean negative mode in the black
hole solution and the classical Gregory-Laflamme instability of the
uncharged black brane. For the case of a black hole in a finite
cavity, the appropriate boundary condition in both the Euclidean
negative mode and the classical perturbation is that the induced
metric on the boundary be unchanged. In the remainder of the paper, we
will show that this Euclidean negative mode (and hence the classical
instability) occurs precisely when the black brane is
thermodynamically unstable.

\section{Schwarzschild in a finite cavity}

Gravitational thermodynamics in a finite cavity was considered by
York~\cite{York:thermodynamics}. He considered four-dimensional
spacetime, with the proper area of the spherical cavity $A = 4\pi
r_b^2$ and the local temperature $T$ at the cavity wall fixed. In the
Euclidean solution, these boundary conditions amount to fixing the
induced metric on the boundary. If the product of cavity radius and
temperature was sufficiently high, $r_b T > \sqrt{27}/8\pi$, there
were two Schwarzschild solutions of mass $M_1, M_2$ which satisfied
the boundary conditions. As $r_bT$ runs from this minimum value to
infinity, $M_1/r_b$ runs from $1/3$ to $0$, and $M_2/r_b$ runs from
$1/3$ to $1/2$ (thus, in the limit, the black hole fills the
cavity). The stability in the canonical ensemble was analysed by
calculating the specific heat at constant area of the boundary, with
the result
\begin{equation} 
C_A=8\pi M^2 
\left(1-\frac{2M}{r_b}\right)\left(\frac{3M}{r_b}-1\right)^{-1}. 
\end{equation} 
Thus, the smaller black hole of mass $M_1$ has negative specific heat
and is thermodynamically unstable, while the larger black hole of mass
$M_2$ has positive specific heat and is thermodynamically stable.

In considering perturbations, it is more convenient to use the black
hole's mass $M$ and $r_b$ as parameters, rather than $T$ and $r_b$,
and we will imagine varying $r_b$. We can consider any $r_b > 2M$. For
$2M < r_b < 3M$, we have a positive specific heat, while for $r_b >
3M$, we have a negative specific heat. We therefore expect to see a
negative mode only for $r_b > 3M$. Allen analysed the problem of
finding the negative mode for black holes in a finite cavity prior to
York's study of the thermodynamics~\cite{Allen:negmode}. He found that
the negative mode exists if $r_b \gtrapprox 2.89M$. A resolution of
this apparent contradiction was presented by
York~\cite{York:thermodynamics}, who observed that Allen had fixed the
coordinate location of the boundary, which, in his choice of gauge for
the perturbation, did not correspond to fixing the area of the
boundary.

In the next section, we will discuss the solution of the eigenvalue
equation (\ref{eqn:eigenvalue}) with the induced metric on the boundary
fixed. First, however, we will briefly discuss the straightforward
extension of York's thermodynamic analysis to higher dimensions. The
$d$-dimensional Schwarzschild black hole is 
\begin{equation} \label{eqn:dschw}
ds^2 = \left( 1 - \frac{\omega M}{r^{d-3}} \right) dt^2 + \left( 1 -
\frac{\omega M}{r^{d-3}} \right)^{-1} dr^2 + r^2 d \Omega_{d-2} , 
\end{equation}
where $\omega = 16 \pi/[(d-2) V_{d-2}]$ and $V_{d-2}$ is the volume of
the unit $S^{d-2}$.  Hence, the relation between mass and temperature
is
\begin{equation}
T(r_b) = \frac{1}{4\pi (\omega M)^{1/(d-3)}} \left( 1 - \frac{\omega
M}{r_b^{d-3}} \right)^{-1/2}, 
\end{equation}
and the specific heat of a black hole in a cavity of radius $r_b$ is 
\begin{equation} \label{eqn:specificheat}
C_A = 4\pi (d-3) M (\omega M)^{1/(d-3)} \left( 1 - \frac{\omega M}{
r_b^{d-3}} \right) \left( \frac{d-1}{2} \frac{\omega M} {r_b^{d-3}}
-1 \right)^{-1}.
\end{equation}
We therefore expect there to be a negative mode in the Euclidean
solution only for $r_b^{d-3} > \frac{d-1}{2} \omega M$.

\section{Finding the Negative Mode} \label{section:calculations}    

We will now discuss the calculation of the negative mode for a
$d$-dimensional Euclidean Schwarzschild black hole in a finite cavity
with the induced metric on the wall of the cavity fixed. By the
foregoing discussion, this also gives a threshold unstable mode for
the uncharged black brane solution. We will first discuss the analysis
for four dimensions in detail and then sketch the extension to higher
dimensions.

\subsection{Four dimensions}

We wish to consider perturbations of the four-dimensional Euclidean
Schwarzschild metric,
\begin{equation} 
ds^2 = \left(1-\frac{2M}{r}\right)d\tau^2 +
\frac{dr^2}{\left(1-\frac{2M}{r}\right)} + r^2d\theta^2
+r^2\sin^2\theta d\phi^2 .
\end{equation}  
We are looking for perturbations which are eigenvectors of
(\ref{eqn:eigenvalue}) with negative eigenvalues. This problem was
analysed in~\cite{Gross:flatspace} for the black hole in infinite
volume, and it was found that in an expansion in spherical harmonics,
the only part which can have a negative eigenvalue is the spherically
symmetric static mode. In considering the black hole in a finite box,
it will therefore suffice for us to consider a perturbation
\begin{equation} \label{eqn:perturbation} 
{h^\mu}_\nu =
Diag\left[H_0(r),H_1(r),-\frac{1}{2}\left(H_0(r)+H_1(r)\right),
-\frac{1}{2}\left(H_0(r)+H_1(r)\right)\right],
\end{equation} 
where the condition of transversality, $\nabla_\mu {h^\mu}_\nu =0$,
implies
\begin{equation} \label{eqn:transversality} 
H_0(r)=\left[-\frac{r(r-2M)}{r-3M}\frac{d}{dr} -
\frac{3r-5M}{r-3M}\right] H_1(r).
\end{equation} 
The task of finding the negative mode then reduces to finding negative
eigenvalues of
\begin{equation} \label{eqn:ode} 
\left[-\left(1-\frac{2M}{r}\right)\frac{d^2}{dr^2} -
\frac{2(r-4M)(2r-3M)}{r^2(r-3M)}\frac{d}{dr} +
\frac{8M}{r^2(r-3M)}\right] H_1(r) = \lambda H_1(r).
\end{equation} 
This equation has regular singular points at $r=0$, $r=2M$ \& $r=3M$
and an irregular singular point at $r=\infty$, so an explicit solution
to  it can not be found.  As in previous works, we must therefore use
numerical methods to find the eigenvalues.  In the infinite cavity the
required boundary conditions for the solution are regularity at the
horizon, $r=2M$, and normalizability at infinity.  Imposing these
conditions on the solution then yields one negative  eigenvalue,
$\lambda \approx -0.19M^{-2}$~\cite{Gross:flatspace}.

To search for negative modes in a finite cavity, we need to impose the
condition that the induced metric on the walls of the cavity is
fixed. This is complicated by the fact that the perturbation is
generically non-zero in the spherical directions.  In the unperturbed
metric the area of a spherical cavity of radius  $r=r_b$ is $A=4\pi
{r_b}^2$.  The perturbed metric is
\begin{equation} 
ds^2=\left(1-\frac{2M}{r}\right)\left(1+ \epsilon H_0(r)\right)d\tau^2
+  \frac{1+ \epsilon H_1(r)}{1-\frac{2M}{r}}dr^2 +
r^2\left[1-\frac{1}{2}\epsilon (H_0(r)+H_1(r))\right]d\Omega^2,
\end{equation} 
and so after the perturbation the surface $r=r_b$ has area
\begin{equation} 
A=4\pi {r_b}^2\left[1-\frac{1}{2}\epsilon (H_0(r)+H_1(r))\right].
\end{equation} 

The change of area could be corrected by allowing the boundary of the
cavity $r_b$ to move to $r'_b$ after the perturbation, where $r'_b$ is
chosen such that the area of the cavity wall in the perturbed metric
is the same as that in the unperturbed metric at $r_b$.  We will
instead give a more careful analysis by making a change of coordinates
to work in a different gauge for the perturbation, in which the area
of the cavity wall at fixed radius is unchanged by the
perturbation. The appropriate coordinate transformation is
\begin{equation} 
r\rightarrow\rho:\quad  \rho^2=r^2\left[1-\frac{1}{2}\epsilon
(H_0(r)+H_1(r))\right].
\end{equation} 
Rewriting the perturbed metric in the new coordinates and keeping only
the dominant first order terms of the metric perturbation gives
\begin{equation} 
ds^2=\left(1-\frac{2M}{\rho}\right) \left(1+ \epsilon
F_0(\rho)\right)d\tau^2 + \left(1-\frac{2M}{\rho}\right)^{-1}
\left(1+\epsilon F_1(\rho)\right)d\rho^2 +\rho^2  d\Omega^2,
\end{equation} 
where
\begin{eqnarray} 
F_0(\rho) & = & \frac{\left(2\rho-3M\right)H_0(\rho)
+MH_1(\rho)}{2\left(\rho-2M\right)}, \nonumber \\ F_1(\rho) & = &
\frac{\left(\rho-3M\right)H_0(\rho) + \left(3\rho-7M\right)H_1(\rho)
+\rho(\rho-2M)\left({H_0}^{\prime}(\rho)
+{H_1}^{\prime}(\rho)\right)}{2\left(\rho-2M\right)}. \nonumber
\end{eqnarray} 
Thus the metric perturbation in this gauge is
\begin{equation}  \label{eqn:newpert}
{h^\mu}_\nu = Diag\left[F_0(\rho),F_1(\rho),0,0\right] .
\end{equation} 
With this new gauge we see that to hold the surface area fixed when
the metric perturbation is switched on is a trivial task: we simply
hold $\rho_b$ fixed.

In the introduction, we saw that physical negative modes are
determined by the eigenvalue equation~(\ref{eqn:eigenvalue}), which
involves only the transverse traceless part of the perturbation. An
infinitesimal coordinate transformation changes the metric by
$g_{\mu\nu} \to g_{\mu\nu} + \epsilon \nabla_{(\mu} \xi_{\nu)}$ for
some vector field $\xi^\mu$, and thus only changes the longitudinal
part. That is, although the metric perturbation in our new gauge
choice looks more complicated, the equation we need to solve is
unchanged, as the transverse traceless part is not affected by the
gauge transformation. The transverse traceless part of
(\ref{eqn:newpert}) is just~(\ref{eqn:perturbation}) with $r \to
\rho$, that is,
\begin{equation} 
{h^{TT\;\mu}}_\nu =
Diag\left[H_0(\rho),H_1(\rho),-\frac{1}{2}\left(H_0(\rho)+H_1(\rho)\right),
-\frac{1}{2}\left(H_0(\rho)+H_1(\rho)\right)\right]
\end{equation}  
with the transversality condition
\begin{equation} \label{eqn:transversalityrho} 
H_0(\rho)=\left[-\frac{\rho(\rho-2M)}{\rho-3M}\frac{d}{d\rho} -
\frac{3\rho-5M}{\rho-3M}\right] H_1(\rho).
\end{equation} 
We are thus seeking negative eigenvalues of the equation
\begin{equation} \label{eqn:oderho} 
\left[-\left(1-\frac{2M}{\rho}\right)\frac{d^2}{d\rho^2} -
\frac{2(\rho-4M)(2\rho-3M)}{\rho^2(\rho-3M)}\frac{d}{d\rho} +
\frac{8M}{\rho^2(\rho-3M)}\right] H_1(\rho) = \lambda H_1(\rho),
\end{equation} 
subject to the boundary conditions of regularity at $\rho = 2M$ and
the isothermal boundary condition at the fixed radial position
$\rho=\rho_b$.  The first boundary condition is imposed by finding a
regular series solution to the differential equation about $\rho =
2M$. There is only one such solution:
\begin{equation} \label{eqn:series} 
H_1(\rho) = \sum_{n=0}^{\infty}a_n{(\rho/M-2)}^n,
\end{equation}
where
\begin{eqnarray*}
a_1 & = & -(\lambda M^2+2)a_0, \\  a_2 & = & \frac{1}{6}(\lambda M^2
+2)(2\lambda M^2 +7)a_0, \\ a_3 & = & -\frac{1}{36}(\lambda
M^2 +5)(2\lambda^2 M^4 +10\lambda M^2 +15)a_0, \\ a_4 & = &
\frac{1}{360} (585+520\lambda M^2+168\lambda^2 M^4 +30\lambda^3 M^6+2
\lambda^4 M^8)a_0.
\end{eqnarray*} 
The isothermal boundary condition fixes the proper length around the
$S^1$ in the $\tau$ direction, which is given by
$\sqrt{g_{\tau\tau}(\rho_b)}\Delta\tau$. It therefore imposes
\begin{eqnarray} 
\left(1-\frac{2M}{\rho_b}\right) & = &
\left(1-\frac{2M}{\rho_b}\right)\left(1+F_0(\rho_b)\right) \nonumber
\\ \Rightarrow F_0(\rho_b) & = & 0 \nonumber \\ \Rightarrow
H_0(\rho_b) & = &  \frac{M}{3M-2\rho_b}H_1(\rho_b). \label{eqn:bdyc}
\end{eqnarray} 
$H_0$ is given in terms of $H_1$ by (\ref{eqn:transversalityrho}), so
the isothermal boundary condition reduces to a mixed boundary
condition for $H_1$,
\begin{equation} \label{eqn:truebc}
\frac{{H_1}^{\prime}(\rho_b)}{H_1(\rho_b)} +
\frac{6{\rho_b}^2-20M\rho_b+18M^2}{\rho_b(\rho_b-2M)(2\rho_b-3M)} =0.
\end{equation} 
This can be contrasted to the condition derived by Allen by holding
$r_b$ fixed~\cite{Allen:negmode},
\begin{equation} \label{eqn:Allenbc}
\frac{{H_1}^{\prime}(r_b)}{H_1(r_b)} + \frac{3r_b-5M}{r_b(r_b-2M)}=0.
\end{equation} 
It should be noted that in deriving these conditions we have
multiplied through by $\rho_b-3M$ in (\ref{eqn:transversalityrho})
[$r_b-3M$ in (\ref{eqn:transversality})].  This is acceptable so long
as $\rho_b \neq 3M$, but for $\rho_b = 3M$, we need to consider more
carefully the true boundary condition (\ref{eqn:bdyc}).

The method used in solving the differential equation is to numerically
integrate from $\rho=2M$ to $\rho=\rho_b$ using the form of the
solution given by~(\ref{eqn:series}) as the initial data.  However,
the numerical method breaks down at $\rho=3M$ due to the singularity
here in the differential equation.  To overcome this problem we find a
power series solution about $\rho=3M$ and find that the solution is in
fact well behaved there.  To evolve our solution through $\rho=3M$ it
is therefore necessary to numerically integrate up to $3M-\delta$ for
some small $\delta$ and use the results of this numerical integration
to fit the power series solution at $3M$ to the data.  This power
series can then be evaluated at $3M+\delta$ and the value  of $H_1$
and its first derivative can be extracted there to provide  new
initial conditions for a numerical solution beginning at  $3M+\delta$
which can then be evolved up to $\rho_b$.

Unlike at $\rho=2M$ where only one of the series solutions was well
behaved, there are two independent well behaved solutions at
$\rho=3M$.  One of these is of order 1 whilst the other is
$O\left((\rho-3M)^3\right)$.  Since the numerical
solution breaks down in a fairly large region around $3M$, we can not
approach the singularity with too small a $\delta$.  It is therefore
necessary to go to fifth order in the series to maintain accuracy with
a $\delta$ of order $10^{-3}$ and so the particular solution to the
differential equation which we require here is a linear combination of
both series. We therefore use the values of $H_1$ and
${H_1}'$ at $3M-\delta$ to find the coefficients $b_0$ and $c_0$ in
the general form for the solution at $\rho=3M$:
\begin{equation}
H_1(\rho) =  \sum_{n=0}^{\infty}\left(b_n + c_n (\rho/M-3)^3\right)(\rho/M-3)^n,
\end{equation}
where
\begin{align*} 
b_1 & = -\frac{4}{3}b_0, & c_1 & = -\frac{11}{6}c_0, \\
b_2 & = \left(\frac{4}{3}+\frac{3}{2}\lambda M^2\right)b_0,  & c_2 &
= \left(\frac{7}{3}-\frac{3}{10}\lambda M^2\right)c_0, \\ b_3 & =
-\left(\frac{40}{27}+5\lambda M^2\right)b_0, &  & \\ b_4 & =
\left(\frac{130}{81} + \frac{15}{2}\lambda M^2 - \frac{9}{8}\lambda^2
M^4\right)b_0, & & \\ b_5 & = -\left(\frac{136}{81} + \frac{79}{9}\lambda
M^2 - 3 \lambda^2 M^4\right)b_0. & & 
\end{align*} 
The eigenvalue spectrum is now found by a shooting method.  An
arbitrary value of $\lambda$ is input, the solution at $\rho=\rho_b$
is found and tested to see if it obeys the correct boundary
condition.  If it doesn't, $\lambda$ is adjusted appropriately and the
process is repeated until $\lambda$ is found to the required
precision.  We have repeated this process with varying cavity sizes so
that the value of the lowest eigenvalue can be plotted against
$\rho_b$.

\begin{figure}[h]   
\begin{center}  
\includegraphics[width=0.5\textwidth,
height=0.18\textheight]{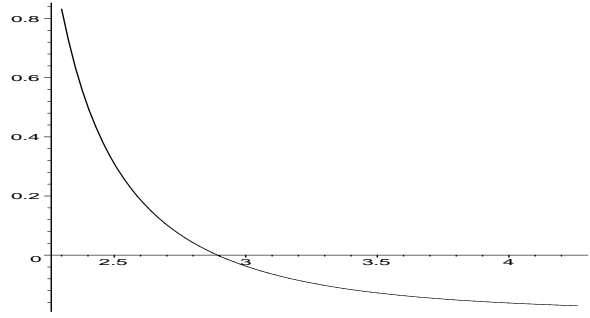}
\end{center}  
\caption{$\lambda M^2$ vs $\rho_b / M$ for the boundary condition
(\ref{eqn:Allenbc})} \label{fig:1}
\end{figure}

\begin{figure}[h]   
\begin{center}  
\includegraphics[width=0.5\textwidth,
height=0.22\textheight]{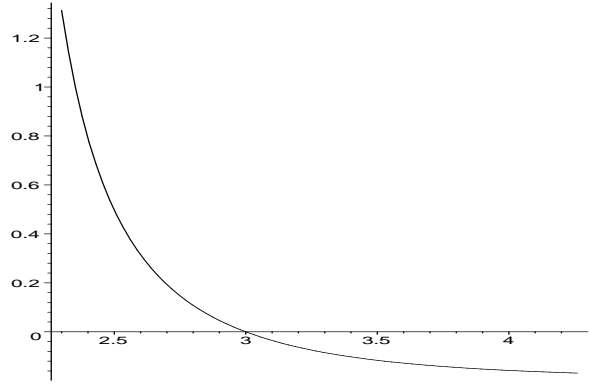}
\end{center}  
\caption{$\lambda M^2$ vs $\rho_b / M$ for the boundary condition
(\ref{eqn:truebc})} \label{fig:2}
\end{figure}

In figure \ref{fig:1}, we give the results obtained for the boundary
condition (\ref{eqn:Allenbc}), demonstrating Allen's  finding that the
critical value of the radius is at $r_b\approx  2.89M$.  In figure
\ref{fig:2}, we give the results for our boundary condition
(\ref{eqn:truebc}), showing clearly the expected result that there is
a negative mode only for $\rho_b > 3M$.

\subsection{Higher dimensions}

\begin{figure}[h]
\begin{center}  
\includegraphics[width=0.85\textwidth, height=0.45\textheight]{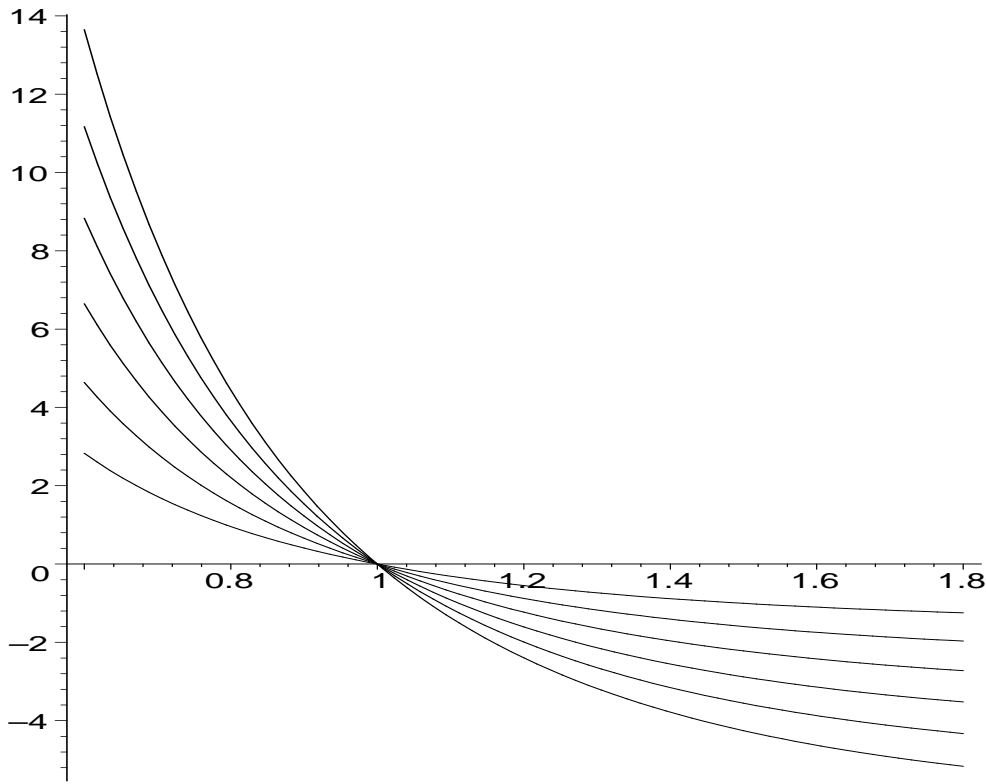}
\end{center}  
\caption{$\lambda (\omega M)^{\frac{2}{d-3}}$ vs $\tilde{\rho}_b$ for
$d=5$ (closest to axis) to $d=10$ (furthest from axis) dimensions}
\label{fig:higherdim}
\end{figure}

We will now briefly discuss the extension of the results above to
uncharged $p$-branes in $d+p$ dimensions for $d>4$. We are interested
in studying the negative modes of the Euclidean black hole geometry
(\ref{eqn:dschw}). In $d>4$, the spherically symmetric transverse
tracefree perturbation is
\begin{equation}
{h^\mu}_\nu =
Diag\left[H_0(r),H_1(r),-\frac{1}{d-2}\left(H_0(r)+H_1(r)\right),
\ldots,   -\frac{1}{d-2}\left(H_0(r)+H_1(r)\right)\right],
\end{equation}
and the transversality condition $\nabla_\mu h^\mu_\nu =0 $ gives
\begin{equation} \label{eqn:dtrans}
H_0(r) = -\frac{2 r (r^{d-3} - \omega M)}{ 2r^{d-3} - (d-1) \omega M}
H_1'(r) - \frac{2(d-1) r^{d-3} - (d+1) \omega M}{ 2r^{d-3} - (d-1)
\omega M} H_1(r).
\end{equation}
The Euclidean negative mode is still given by the eigenvectors of
(\ref{eqn:eigenvalue}), which becomes
\begin{eqnarray} \nonumber
&-&\left( 1 - \frac{\omega M}{ r^{d-3}} \right) H''_1(r) - \frac{[2d
r^{2d-6} - \omega M r^{d-3} (3d^2 -11d+18) + d(d-1) \omega^2 M^2]}{
r^{d-2} (2r^{d-3} - (d-1) \omega M)} H'_1(r) \\&&+ \frac{2d(d-3)^2
\omega M}{ r^2 (2r^{d-3} - (d-1) \omega M)} H_1 (r) = \lambda
H_1(r).
\end{eqnarray}
As in the four-dimensional case, it is convenient to make a coordinate
transformation before applying the boundary conditions at the
shell. Here the appropriate transformation is
\begin{equation}
r\rightarrow\rho:\quad  \rho^2=r^2\left[1-\frac{1}{d-2}\epsilon
(H_0(r)+H_1(r))\right].
\end{equation}
The condition that the induced metric on the shell is fixed then
implies that the shell is at some fixed $\rho = \rho_b$, and that
\begin{equation}
H_0(\rho_b) + \frac{d-3}{ 2 (d-2)} \frac{\omega M}{ \rho_b^{d-3} -
 \omega M} (H_0(\rho_b) + H_1(\rho_b)) = 0.
\end{equation}
Using the transversality condition (\ref{eqn:dtrans}), this becomes a
mixed boundary condition for $H_1$,
\begin{equation}
\frac{H_1'(\rho_b)}{ H_1(\rho_b)} + \frac{2(d-1)(d-2) \rho_b^{2d-6} -
2(d^2-d-2) \omega M \rho_b^{d -3} + (d-1)^2 \omega^2 M^2}{ \rho_b
(\rho_b^{d-3} - \omega M) [2(d-2) \rho_b^{d-3} - (d-1) \omega M]} = 0.
\end{equation}
We analyse this eigenvalue problem using the same numerical methods as
previously. The results are displayed in figure~\ref{fig:higherdim}.
In this graph the eigenvalue has been plotted against the cavity
radius in a scaled and shifted radial coordinate, $\tilde{\rho}$, given by
\begin{equation}
\tilde{\rho} = \frac{\rho(\omega
M)^{\frac{1}{3-d}}-1}{\left(\frac{d-1}{2}\right)^{\frac{1}{d-3}}-1}.
\end{equation}
This choice is made so that in all dimensions we see that the black
hole horizon is at $\tilde{\rho}=0$ and the specific
heat~(\ref{eqn:specificheat}) is negative only when
$\tilde{\rho}_b>1$.  It is clear that in all cases this corresponds
exactly to the existence of the negative mode.

\section{Conclusions}  

We have demonstrated that an uncharged black brane in a spherical
cavity is classically unstable if and only if it is locally
thermodynamically unstable. This provides a particularly simple and
elegant example of the general connection between thermodynamic and
classical instability conjectured
in~\cite{Gubser:instability1,Gubser:instability2}.  The key element of
the proof was showing that as we vary the volume of the cavity, the
classical instability disappears at precisely the point where the
specific heat changes sign. To do this, we used the observation that
the threshold unstable mode for the classical instability is the
analytic continuation of the Euclidean negative
mode~\cite{Reall:branes}. We then provided the first analysis of the
Euclidean negative mode for Schwarzschild in a finite cavity using the
boundary conditions appropriate to the canonical ensemble. (A similar
calculation of the negative mode for the Euclidean Schwarzschild
anti-de Sitter solution was carried out in~\cite{Prestidge:stability};
unfortunately, it is not straightforward to construct a corresponding
black brane solution in that case.  The issue of stability for a warped
black string solution in anti-de Sitter space was considered in~\cite{Hirayama:blackstrings}.)

One might be surprised that the presence of a boundary affects the
classical instability at all; after all, the bulk metric is unchanged,
so the initial behaviour of a perturbation with compact support near
the horizon should be unaffected by the introduction of the
boundary. However, in general relativity, the initial data are subject
to constraints, so we are not free to specify an arbitrary initial
perturbation of compact support, and the boundary conditions can
influence the allowed possibilities for initial perturbations. It is
hence reasonable that the boundary can turn off the classical
instability.

\vskip.5in

\centerline{\bf Acknowledgements}
\medskip    
    
We are grateful for discussions with Dominic Brecher, Lee Garland, Ken
Lovis and Paul Saffin.  The work of JPG is supported in part by EPSRC
studentship 9980045X.

\bibliographystyle{utphys}  
   
\bibliography{NegativeMode}   
    
\end{document}